\documentclass[sn-mathphys-num]{sn-jnl}% Math and Physical Sciences Numbered Reference Style 
\usepackage{graphicx}%
\usepackage{multirow}%
\usepackage{amsmath,amssymb,amsfonts}%
\usepackage{amsthm}%
\usepackage{mathrsfs}%
\usepackage[title]{appendix}%
\usepackage{xcolor}%
\usepackage{textcomp}%
\usepackage{manyfoot}%
\usepackage{booktabs}%
\usepackage{algorithm}%
\usepackage{algorithmic}%
\usepackage{listings}%
\usepackage{comment}
\usepackage{caption}
\usepackage{xcolor}
\usepackage{url}
\usepackage{subcaption}

\theoremstyle{thmstyleone}%
\theoremstyle{thmstyletwo}%

\theoremstyle{thmstylethree}%

\begin{document}

\title[Article Title]{GPS Spoofing Attacks on AI-based Navigation Systems with Obstacle Avoidance in UAV}

%%=============================================================%%
%% GivenName	-> \fnm{Joergen W.}
%% Particle	-> \spfx{van der} -> surname prefix
%% FamilyName	-> \sur{Ploeg}
%% Suffix	-> \sfx{IV}
%% \author*[1,2]{\fnm{Joergen W.} \spfx{van der} \sur{Ploeg} 
%%  \sfx{IV}}\email{iauthor@gmail.com}
%%=============================================================%%

\author[1]{\sur{Ji Hyuk Jung}}\email{graycat@korea.ac.kr}

\author[1]{\sur{Mi Yeon Hong}}\email{hachikohmy@korea.ac.kr}

\author[1]{ \sur{Ji Won Yoon}}\email{jiwon\_yoon@korea.ac.kr}

\affil[1]{\orgdiv{School of CyberSecurity}, \orgname{Korea University}, \orgaddress{\street{Anam-ro}, \city{Seoul}, \postcode{02841}, \state{Seoul}, \country{Korea}}}

%%==================================%%
%% Sample for unstructured abstract %%
%%==================================%%

\abstract{Recently, approaches using Deep Reinforcement Learning (DRL) have been proposed to solve UAV navigation systems in complex and unknown environments. However, despite extensive research and attention, systematic studies on various security aspects have not yet been conducted. Therefore, in this paper, we conduct research on security vulnerabilities in DRL-based navigation systems, particularly focusing on GPS spoofing attacks against the system. Many recent basic DRL-based navigation systems fundamentally share an efficient structure. This paper presents an attack model that operates through GPS spoofing attacks briefly modeling the range of spoofing attack against EKF sensor fusion of PX4 autopilot, and combine this with the DRL-based system to design attack scenarios that are closer to reality. Finally, this paper experimentally demonstrated that attacks are possible both in the basic DRL system and in attack models combining the DRL system with PX4 autopilot system.
}


%%\pacs[JEL Classification]{D8, H51}

%%\pacs[MSC Classification]{35A01, 65L10, 65L12, 65L20, 65L70}

\maketitle

\section{Introduction}
\label{intro}
Unmanned Aerial Vehicles (UAVs) refer to vehicles that can fly autonomously or be piloted remotely. UAVs are being widely utilized across various domains, particularly for delivery services, surveillance operations, reconnaissance missions in hazardous regions, and agricultural monitoring \cite{chmaj2015distributed, restas2015drone, busby2019drone, jung2017analysis,paucar2018use}. Additionally, intelligent UAV on the battlefield has become a key point in the defense industry \cite{luckey2022anduril, dew2024us}. The development of precise sensors such as GPS and IMU, the advancement of propulsion technologies like motors and engines, and the evolution of estimation or control software made the birth of UAVs possible. Recently, with the rapid advancement of artificial intelligence and the increasing sophistication of UAV applications, UAVs are now expected to perform far more complex flight maneuvers that go well beyond basic automatic flight control. Especially, UAVs are required to perform autonomous flight in complex environments with numerous obstacles, such as buildings in urban settings and natural terrain features in hazardous areas. Recently, vision-based navigation, which performs navigation functions in complex environments based on depth images, has been gaining attention. Traditional autonomous operation either lacks obstacle avoidance capabilities or requires prior knowledge of the shape and position of surrounding obstacles \cite{chen2022end, mac2016improved, escobar2018r}. However, DRL-based navigation systems allow UAVs to operate in complex environments without any prior information. Therefore, DRL-based navigation systems are becoming central to research on autonomous UAV flight. DRL-based navigation system for UAVs primarily utilize depth images, position and direction as fundamental input parameters. DRL-based navigation systems are designed to perform two critical functions: obstacle avoidance and target-oriented navigation. Recent research in the field of DRL-based navigation systems has converged on an efficient architectural paradigm, which typically processes depth images through CNN layers, followed by a concatenation layer where physical state inputs such as position and direction are combined.

Recently, research on attacks and defenses related to adversarial attacks on these DRL systems has been conducted \cite{hickling2023robust}. While this demonstrates growing security concerns about these systems, there has been little research conducted beyond this area so far. Fault injection or sensor spoofing attack on robotic vehicles have been major topics of interest in security research for a long time \cite{sato2021dirty, zhong2022neural, zhou2022strategic, quinonez2020savior, shen2020drift, khan2024comprehensive}. In \cite{zhou2022strategic}, a attack model and methodology were proposed to cause malfunctions in ADAS systems through fault injection, assuming the attacker can access internal or external network systems. A fuzzing-based attack was proposed under the assumption that the attacker can input data to all sensor inputs, including image sensors and LiDAR sensors in \cite{zhong2022neural}. \cite{sato2021dirty} demonstrated that an adversarial attack for autonomous vehicle can be conducted in a physical environment. While numerous attacks on sensors and inputs have been conducted against autonomous vehicles, attack modeling for DRL-based navigation systems has not yet been researched. Therefore, from this perspective, this paper aimed to design an attack model for sensor spoofing attacks on DRL-based systems and experimentally prove that such attacks are feasible. This paper proposes GPS spoofing attack scenario that is feasible within DRL-based navigation systems. While research has been conducted on adversarial attacks against depth images in DRL-based systems \cite{hickling2023robust}, research on sensor spoofing attacks targeting input values for position and direction have not yet been carried out. Additionally, adversarial attacks based on depth images have limitations regarding their practicality in real-world environments because the high dimensional nature of depth images, characterized by a large number of pixels, necessitates subtle and precise alterations to individual pixel values for effective adversarial attack. Especially, there has been little research on GPS spoofing attack for the DRL-based navigation system. Therefore, this paper analyzes the architecture of the DRL-based system for GPS spoofing attack and systematically proposes attack methodologies. To consider GPS spoofing detection, this study investigates the feasible range of spoofing attacks from the perspective of UAV autopilot system and restricts the input values which avoid detection by the spoofing detection system in the DRL-based system model. In summary, this paper makes the following contributions.
\begin{itemize}
    \item This paper proposes threat scenarios and analyses the recently studied DRL-based system for modeling GPS spoofing attacks.
    
    \item To perform GPS spoofing attack on recent UAV autopilot controller systems, attackers must evade the spoofing detection mechanisms of the control system, which limits the attacker's input space. This paper integrates these constraints to present a comprehensive threat attack model.
    
    \item The paper experimentally demonstrates that GPS spoofing attack on the DRL-based navigation system are feasible even within this constrained input space.
\end{itemize}

\begin{comment}

즉, 본 논문은 첫째로 최근 연구되고 있는 DRL-based UAV에 대한 GPS spoofing 공격에 관한 위협 시나리오 및 모델을 제안하였으며, 이를 위해 공격이 가능한 인풋 스페이스에 대한 분석을 진행하였다. 둘째로는

최근 UAV 시스템에서의 GPS Spoofing을 하기 위해서는 오토파일럿 컨트롤 시스템의 스푸핑 감지를 회피해야하며, 이는 공격자의 인풋스페이스를 제한한다. 본 논문은 이를 통합아여 위협 공격 모델을 제시하였다. 마지막으로는 이러한 제한된 인풋스페이스에서 DRL 시스템에 대한 공격이 가능함을 실험적으로 보여주었다.

이로써, 본 논문에서는 DRL 시스템에 대한
 realistic threat to DRL-based system.

Contri를 명확하게, 디스커션에 REAL에서의 환경에 대해서는 시뮬이 DRL이 오히려
완고한 시스템이며, 현실에서는 시뮬보다 불안정함. 하지만, 스푸핑 공격의 한계점도 있음.

graycat:
This approach demonstrates that GPS spoofing attack represents a more realistic threat to DRL-based system. Additionally, this research shows that it is possible to attack artificial intelligence systems while taking into account anomaly detection module.
\end{comment}

\section{Background}
\label{sec:Back}

\subsection{Deep Reinforcement Learning}
Deep Reinforcement Learning (DRL) is an advanced paradigm that integrates deep learning techniques with traditional reinforcement learning frameworks \cite{li2017deep}. This synergistic approach enables the utilization of high dimensional data, such as images, as state representations or observations within the reinforcement learning process. The fundamental approach in RL starts with an intelligent agent interacting with a given dynamic environment at each time step. At each time step t, the agent is at the current state $s_{t}$ decides on an action $a_{t}$, and in return, receives its next state $s_{t+1}$ and receives a reward $r(s_{t}, a_{t})$ from the environment. The ultimate goal of reinforcement learning is to maximize the cumulative sum of rewards G, which accumulates until the end of the state sequence. $\gamma$ is a factor used for discounting future rewards. Equation \ref{eq:cumulative} represents the accumulated discounted rewards from the current time step t. The goal of reinforcement learning is to find a policy that maximizes G while the agent interacts with the environment.

\begin{equation}\label{eq:cumulative}
\begin{aligned}
G_{t} = \sum_{t=0}^T\gamma^{t}r(s_{t}, a_{t}).
\end{aligned}
\end{equation}

A policy is a function that outputs the action to be taken based on the current state. The following represents the policy function.
\begin{equation}
\begin{aligned}
\pi_{\theta}(s) = a,
\end{aligned}
\end{equation}
where $\theta$ is the parameter of the policy $\pi$, $s$ and $a$ are the state and action of an agent, repectively. Fundamentally, DRL proceeds with learning to optimize the policy function from sequences of states and actions obtained by the agent in the environment. This sequence is termed a trajectory $\tau$, and is represented by the following equation.

\begin{equation}
\begin{aligned}
\tau = (s_{0}, a_{0}, s_{1}, a_{1}, ... , s_{T}, a_{T}.
)
\end{aligned}
\end{equation}

Policy-based reinforcement learning aims to maximize the expected value of the cumulative reward G from these obtained trajectories. Therefore, the learning process can be represented as follows.

\begin{equation}
\begin{aligned}
J(\theta) = E_{\tau \sim p_{\theta}(\tau)}[\sum_{t=0}^T\gamma^{t}r(s_{t}, a_{t})] \\
\theta^* = \arg \max_{\theta} J(\theta)
\end{aligned}
\end{equation}

As the target of attack for this paper, we selected a navigation system implemented with Twin Delayed DDPG (TD3), one of the most recent DRL methodologies \cite{he2020deep}. TD3 is essentially a DRL algorithm based on Deep Deterministic Policy Gradient (DDPG) \cite{lillicrap2015continuous}, but it addresses several of its shortcomings. TD3 is an off-policy actor-critic algorithm which learns a policy and Q function simultaneously . Additionally, TD3 assumes a policy function to be a deterministic. Q function outputs the expected value when a specific action is taken in a given state. For learning Q function, TD3 uses replay buffer technique, which involves randomly extracting states, actions, and rewards from trajectories to facilitate learning \cite{mnih2013playing}. Equation \ref{eq:Q} represents the loss function used for learning the Q function.

\begin{equation}\label{eq:Q}
\begin{aligned}
L(\phi) = E_{(s,a,r,s') \sim D}[(Q_{\phi} - (r+ \gamma Q_{\phi}(s',\pi_{\theta}(s'))))^2].
\end{aligned}
\end{equation}

Modeling Q function and assuming a deterministic policy function simplify the calculation of the gradient of the objective function $J(\theta)$.

\begin{equation}
\begin{aligned}
\nabla_{\theta} J(\theta) = \nabla_{\theta} \sum_{k=0}^N Q_{\phi}(s_{k}, \pi_{\theta}(s_{k}))
\end{aligned}
\end{equation}

\begin{equation}
\begin{aligned}
\phi' \leftarrow \tau \phi + (1- \tau) (\phi + \alpha \nabla_{\phi} L(\phi)) \\
\theta' \leftarrow \tau \theta + (1- \tau) (\theta + \alpha \nabla_{\theta} L(\theta))
\end{aligned}
\end{equation}

TD3 introduces three distinctive methodologies to address the shortcomings of deterministic policy-based methods, which include target policy smoothing, clipped double-Q learning, and delayed policy updates \cite{fujimoto2018addressing}. The deterministic policy in DDPG simplifies calculations but can lead to overfitting. To mitigate this issue, the algorithm introduces clipped double-Q learning which uses two action value functions. Each Q function is updated separately, but to prevent overfitting, the TD target is shared using the smaller output value of the two Q functions during updates. Target policy smoothing enhances policy stability and facilitating exploration by adding noise into the policy actions and clipping the action. Delayed Policy Updates is the approch that updates the policy only after the critic has converged to a certain degree rather than updating the policy and critic at every step.

\subsection{AI for UAV Navigation with Obstacle Avoidance}\label{sub:DRL_NAV}

The conditions and goals of DRL-based navigation system are as follows:
    \begin{itemize}
    \item The system possesses knowledge solely of its target destination, lacking any prior information about the surrounding environment, including the location and configuration of obstacles.

    \item While navigating towards its designated target position, the UAV must demonstrate the capability to effectively circumvent obstacles.

    \item Utilizing depth image, relative position, and direction, the system autonomously determines and executes appropriate control actions.
    \end{itemize}

Numerous studies have focused on developing the most efficient system that satisfies these functionalities and conditions \cite{he2020deep, kalidas2023deep, he2021explainable, krishnan2021air, hickling2023robust, cetin2019drone, zhang2022autonomous}. Recent studies predominantly share a common structure for neural network that outputs actions, fundamentally considering both functionality and learning efficiency. This structure involves depth image input being processed through convolution layers for feature extraction, after which the UAV's physical states such as relative distance and direction are combined with the extracted image feature in a concatenation layer. Subsequently, in the concatenation layer, the combined information passes through multiple linear layers to ultimately output the action. The physical state has a significantly smaller dimension compared to the depth image. Therefore, processing it separately is more efficient, as combining it with the depth image during training may result in poor reflection of this information.

\begin{figure}[H] 
\centering
\includegraphics[width=8cm, height=4cm]{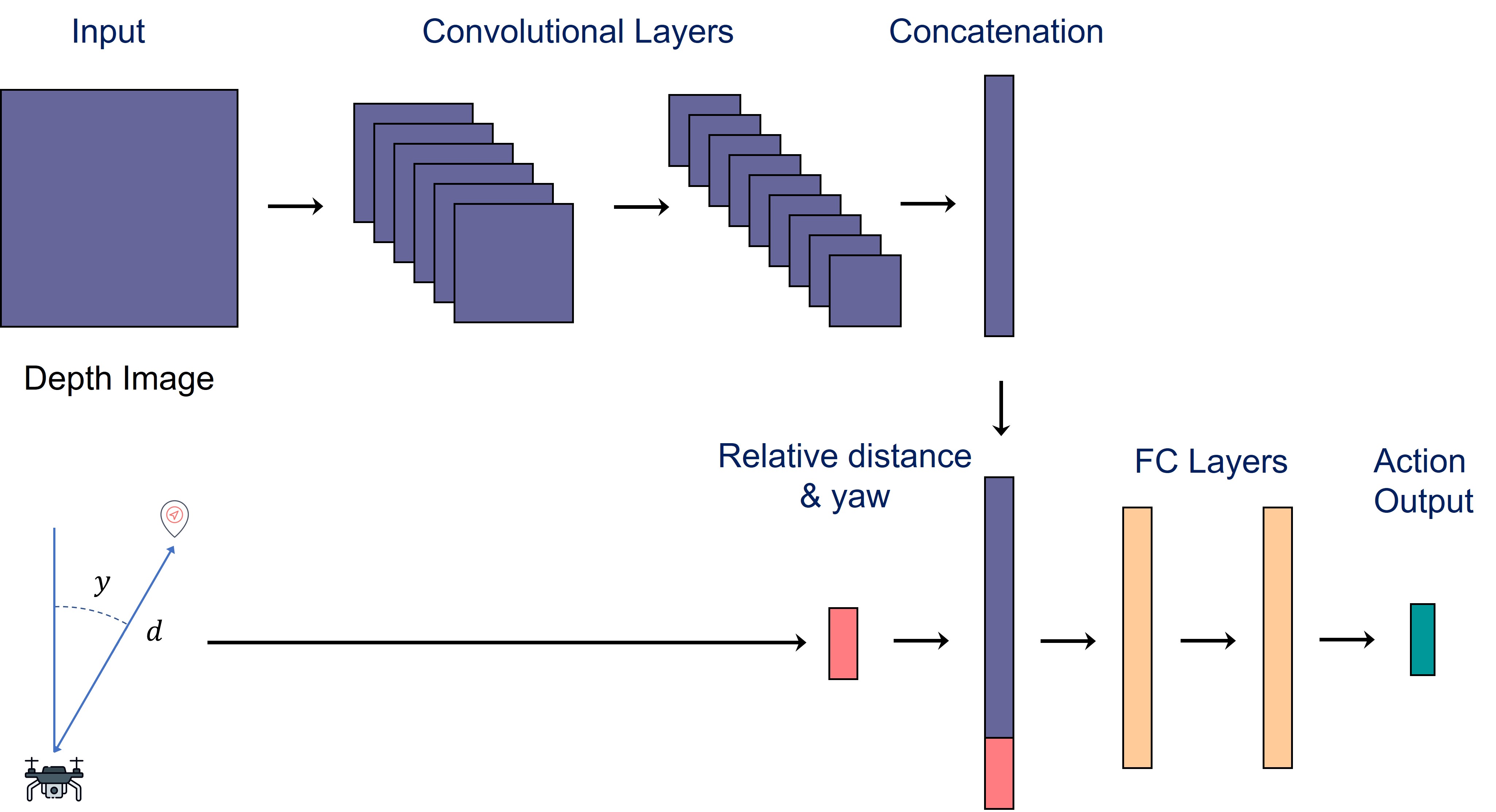}
\caption{The basic structure commonly shared by recent studies when designing navigation systems based on DRL.}\label{fig:shared_model}
\end{figure} 

\subsection{Sensor Fusion and Anomaly Detection}\label{subsec:sensor_fusion}
Extended Kalman Filter (EKF) is an algorithm used to estimate the state of a system from sensor measurements and a system's nonlinear dynamic model \cite{kalman1960new}. EKF sensor fusion combines sensor measurements collected from multiple sensors and estimates the system state, which is achieved through two main processes. One process involves updating the system state through the dynamic model, during which sensor inputs are not specifically incorporated. Another process involves incorporating sensor measurements into the system state. This process reflects the difference between the internal estimates of the system and the values estimated from the sensors. The latest robotic vehicle systems detect a significant discrepancy, it does not incorporate this difference and instead determine it as an anomaly value \cite{px4, liu2019analysis, quinonez2020savior, noh2019tractor, jung2024analysis}. If sensor measurements are continuously identified as anomalies, they may be classified as a spoofing attack, leading to the suspension of sensor data incorporation in robotic vehicles.

The process of incorporating sensor values in EKF sensor fusion can be summarized as follows:
\begin{equation}
\begin{aligned}
est_{next}' \leftarrow pred(est_{cur}) \\
est_{next} \leftarrow est_{next}' + K \times (obs_{cur} - h(est_{next}'))
\end{aligned}
\end{equation}
where $est_{next}'$ represents the values predicted from the dynamic model based on the system's internal estimates $est_{cur}$ before incorporating the sensor values, $K$ indicates the degree to which sensor values are incorporated. $est_{next}'$ is corrected through the $obs_{cur}$ obtained from the sensor and becomes the next system internal state $est_{next}$. $h$ is the relationship function between the sensor values and the internal state estimates. Generally, for each update of the internal system estimates through sensor values, the system decides whether to incorporate the sensor values based on the magnitude of $(obs_{cur} - h(est_{next}'))$  and this is also used for anomaly detection. This paper models the practical range of possible spoofing to assume more realistic attacks in PX4 which is the most widely used UAV controller systems in research and industry \cite{px4}. The incorporation of sensor values and anomaly detection is fundamentally applied in PX4. In PX4, the test ratio is calculated using $(est_{next}' - obs_{cur})$ to detect anomalies. In other words, if the test ratio exceeds 1, the sensor values are not incorporated, and prolonged non-incorporation is used to detect spoofing attacks or abnormal sensor states.

\section{Proposed Method}

\subsection{Attack Target System Model}\label{subsec:TargetModel}
This paper proposes modeling an attack target system that extends from the anomaly detection and sensor estimation algorithm layer of PX4 to the DRL navigation model. As mentioned in Section \ref{sub:DRL_NAV}, the DRL-based navigation system outputs the velocity that a UAV should take through inputs of depth image, relative position, and yaw. Basically, in the DRL system, the target position is determined at the beginning of a mission, and the current position is continuously estimated in EKF module in PX4, allowing to calculate the relative position to the target. The relative direction is determined by calculating the difference between the current direction of the UAV and the direction to the target position, based on the current UAV position which is an estimated value from PX4. The relative distance and direction, which are inputs to the DRL system, are calculated as follows, with the direction expressed as the angle yaw.

\begin{equation}\label{eq:realtive}
\begin{aligned}
d_{rel} &= \|{pos_{tar} - pos_{cur}}\| \\
yaw_{tar} &= \arctan(pos_{tar}-pos_{cur}) \\
yaw_{rel} &= yaw_{tar} - yaw_{cur} \\
\end{aligned}
\end{equation}
The relative distance and yaw are denoted by $d_{rel}$ and $yaw_{rel}$, respectively. Meanwhile, $pos_{tar}$ represents the target position, and $pos_{cur}$ signifies the current position. $yaw_{rel}$ is the angular direction of the target relative to the current position. In particular, $pos_{cur}$ refers to the current position state that the UAV recognizes, which corresponds to the system internal state mentioned in subsection \ref{subsec:sensor_fusion}.

\begin{figure}[H] 
\centering
\includegraphics[width=8cm, height=4cm]{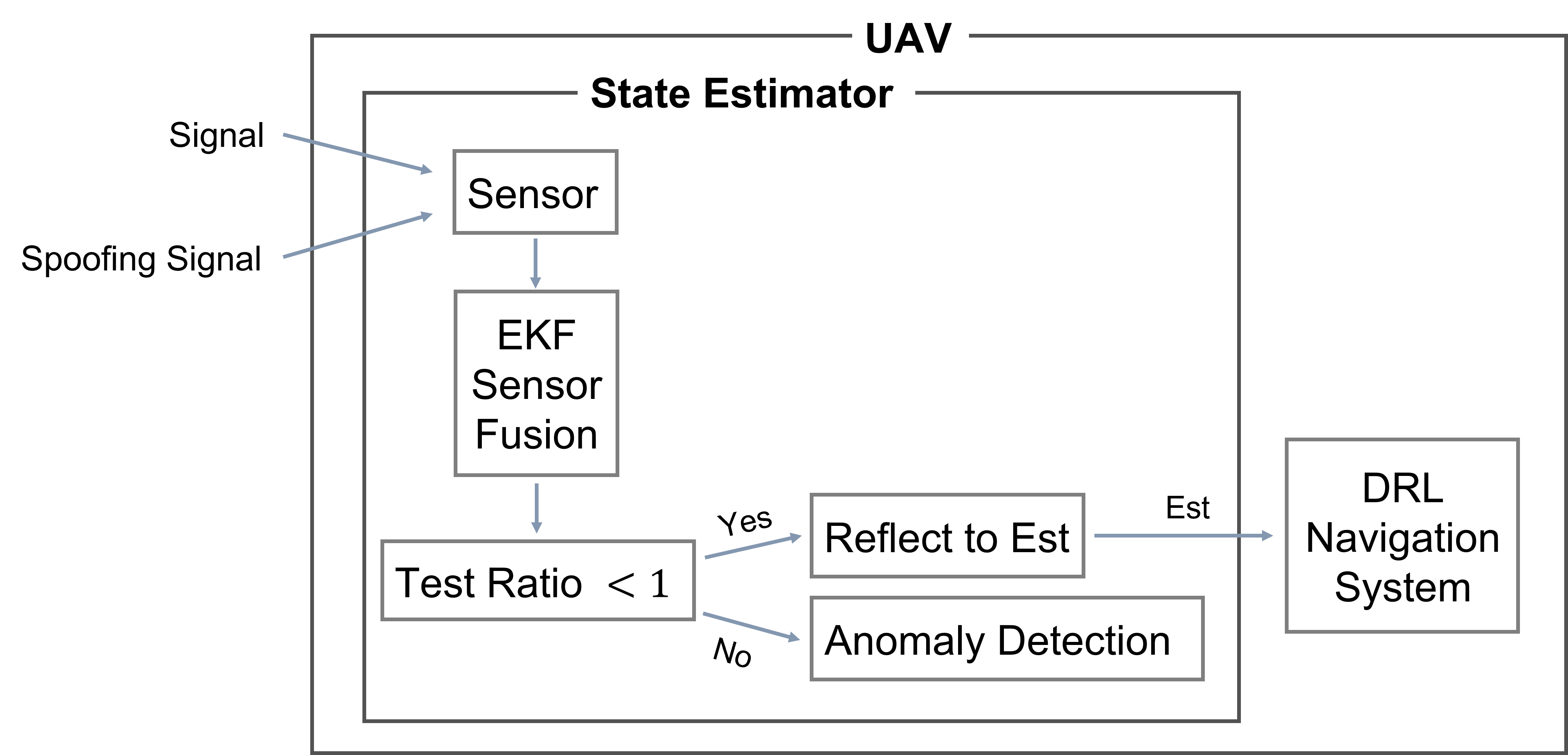}
\caption{The practical target model models the spoofing attacks constrained by PX4, and these processed values in PX4 are then passed to the DRL-based navigation system.\label{fig:target_model}}
\end{figure}

\subsection{Base Principle of Attack}\label{subsec:principle}

The DRL-based system for UAV collision avoidance and navigation to a specific target, which is currently used in most research, basically has the following structure as shown in Figure \ref{fig:shared_model}. The relative direction to the target alone is insufficient to distinguish between scenarios where the target is positioned between the obstacle and the UAV, and where the target is behind both the UAV and the obstacle. Therefore, the relative distance of the target becomes an essential input value. However, this structure has vulnerabilities. The first issue arises because the the decision by a depth image contradicts the decision based on physical states such as UAV's direction or its relative position to the UAV's target. Fundamentally, the DRL system processes a depth image and relative physical states of the target simultaneously in the same neural network for handling both movement towards the target and movement to avoid collisions at the same time. Even in situations where the distribution of obstacles around the UAV is the same, the distance and direction to target can lead to different actions. This means that the action output can vary depending on the relative physical state with the target, even when obstacles are very close. 

Secondly, physical states such as relative distance, and direction are directly input into the concatenation layer of DRL, thus having a relatively large impact on the action output. While a depth image are difficult to spoof, physical states of UAVs are already relatively easy to manipulate through spoofing attacks like GPS spoofing and electromagnetic interference. As a result, the DRL system is inherently vulnerable to spoofing attacks.

Finally, because a UAV controller must inevitably allow for some degree of sensor error, spoofing within a small range remains possible. The most commonly used UAV autopilot systems like PX4 tend to accept a certain range of error values and reflect them as shown in subsection \ref{subsec:sensor_fusion}. Furthermore, unlike adversarial attacks on depth images, it is known that these spoofing attacks are easily feasible in reality. In particular, systematic methodologies for GPS spoofing attacks are well-researched \cite{noh2019tractor, feng2018efficient, tippenhauer2011requirements, sathaye2022experimental, jung2024analysis}. Additionally, methods for sensor disruption using magnetic fields or ultrasound have also been studied \cite{chen2023magnetic, kim2024systematic, jang2023paralyzing}.

\subsection{Attack Scenario}

\begin{figure}[H] 
\centering
\includegraphics[width=6.6cm, height=4.2cm]{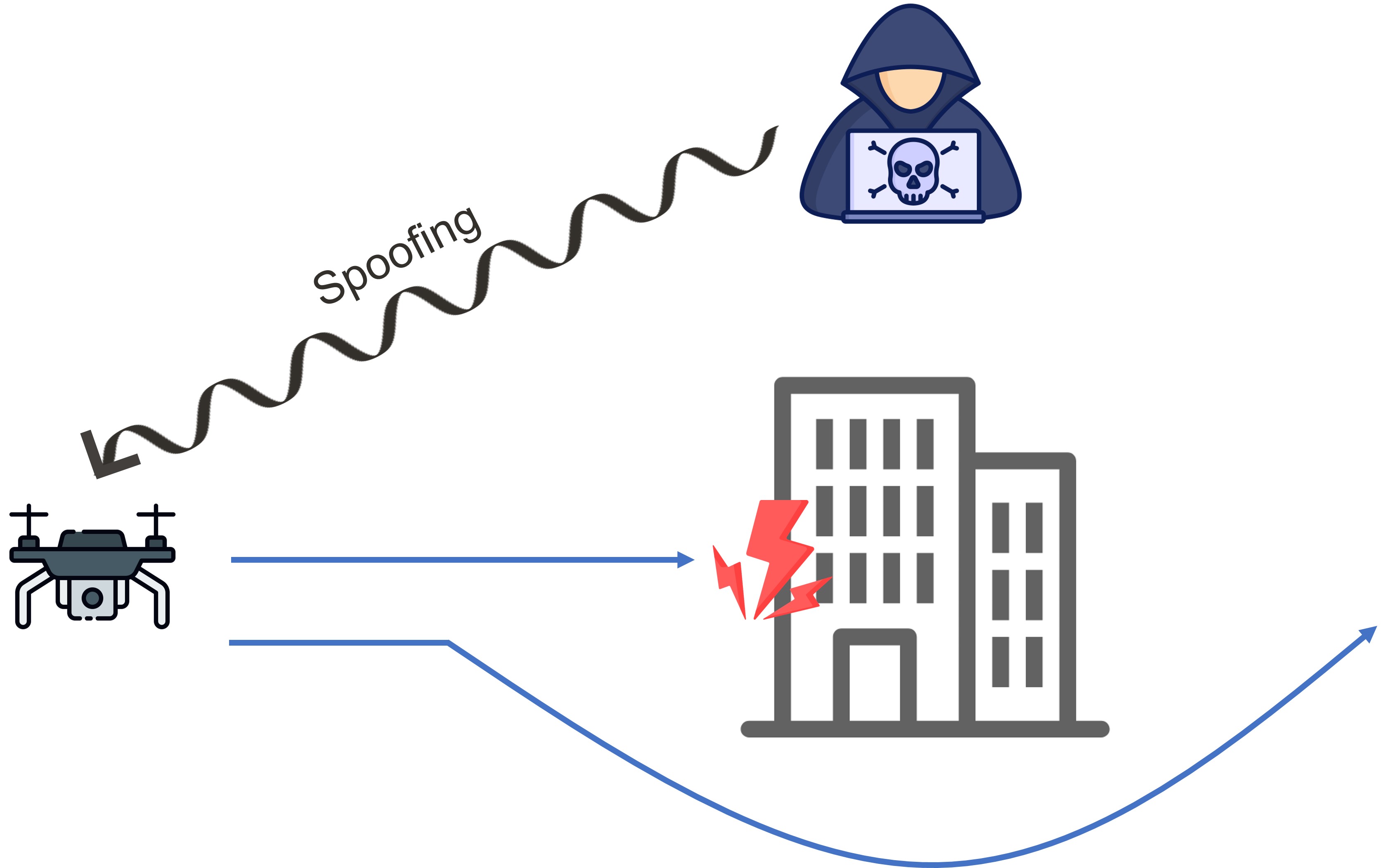}
\caption{A scenario where an attacker induces collisions through spoofing on a UAV system navigating in complex environments such as urban, battlefield or natural disaster areas.\label{fig:attack_model}}
\end{figure}

In modern battlefields and complex urban environments, the operation of intelligent UAVs is required to be possible even in extremely complex geographical settings. The goal of the attacker in this attack scenario is to neutralize and disrupt intelligent UAVs by inducing collisions with obstacles. In other words, the core of the attack scenario in this paper is to interfere with obstacle avoidance, which is a core function of intelligent UAVs, through attack vector inputs related to the physical state. This paper assumes a scenario where an attacker performs GPS spoofing attacks on a UAV with the DRL-based system. Figure \ref{fig:attack_model} illustrates the attack scenario presented in this paper.

\subsection{Unconstrained Attack Model}\label{sub:General}
The attack model in this subsection assumes that there are no limitations on the range of attack vectors used as GPS spoofing input to the DRL system. It assumes a situation where the anomaly detection function for sensor inputs in the UAV's control system is not activated. Even if the range is exceeded, it is not detected by autopilot control system. In other words, this can be seen as an attack model for targets that does not consider EKF sensor fusion detection layer, and it can be viewed as an attack model equal to the models used in general adversarial attacks \cite{hickling2023robust}.
 
This paper proposes a brute force attack methodology to induce collisions in this attack model. In a UAV near collision objects, the attacker seeks to find values in the input space of relative position and yaw that produce outputs guiding the UAV towards a collision. Subsequently, the attacker can inject the values as relative position and yaw values into the DRL system. Algorithm \ref{alg:brute} represents our brute force attack process. $pos$ is the spoofed position inputs of the DRL. An attacker examines the output values for all inputs up to $P$, which are the maximum input values for $pos$ to find the value that meets the condition which induces a collision. The current position of the UAV  can determine the relative distance $d_{rel}$ and yaw $y_{rel}$ to the target, which are inputs to the policy algorithm $\pi$ of the DRL, through Equation \ref{eq:realtive}. $Condition$ refers to a function that, from the attacker's perspective, determines whether the policy's output $\pi(s)$ can induce a collision. In other words, the desired relative distance and yaw can be determined by spoofing the current position. Finally, these values are used for a spoofing attack on the DRL system. Additionally, it is assumed that the attacker knows the input of the depth image. In the attack scenario, it is assumed that the depth input that the UAV is looking at can be sufficiently estimated through the UAV's position and orientation.

\begin{algorithm}
\caption{Unconstrained Attack Model}\label{alg:brute}
\begin{algorithmic}
\STATE $UAV Mission$
\WHILE{$!done$}
\FOR{$pos=0$ to $P$}
\STATE $d_{rel} \gets \sqrt{pos_{tar}-pos)}$
\STATE $y_{rel} \gets \arctan(pos_{tar}-pos) - yaw_{cur}$
\STATE $s \gets (Image, d_{rel}, y_{rel})$
\IF{$Condition(\pi(s))$ == true}
    \STATE $attacker \ transmit \, pos$
\ENDIF
\STATE $UAV action$
\ENDFOR
\ENDWHILE
\end{algorithmic}
\end{algorithm}

\subsection{Constrained Attack Model}\label{sub:practical}

Currently, UAVs typically have built-in anomaly detection capabilities, which prevent the incorporation of detected anomalous sensor values as described in subsection \ref{subsec:sensor_fusion}. Therefore, this paper establishes a more practical attacker model, targeting the system described in subsection \ref{subsec:TargetModel} and presenting a more realistic attack threat.
The conditions of this attack model are as follows:
    \begin{itemize}

    \item The attacker can only interfere with the sensor inputs of the target system described in subsection \ref{subsec:TargetModel}.

    \item The attacker does not know the internal state of the controller system.

    \item The attacker operates in a way that can evade anomaly detection of the controller system.
    \end{itemize}

To overcome this, we proposes methods for spoofing sensor values and explores the extent to which such spoofing is possible. Through this analysis, we have designed a practical attack model.

\begin{comment}

내 논문 인용해야함.

섹션 ?에서 말했듯이, UAV의 내부 위치추정값에서 벗어나게 스푸핑 공격을 하면, 이상탐지가 되기 때문에 공격은 실패한다. 따라서, 공격자는 UAV의 내부 위치추정값내에서 스푸핑값을 주어야한다. 하지만, 

현재로써 탐지기능을 회피하기 위해 기본적으로 스푸핑을 위해서는 최소 공격자는 UAV의 실제위치는 관찰가능하다는 전제가 필요하다.  
\end{comment}

\textbf{Spoofing Strategy in PX4}: 
Despite the fact that the internal estimated values of sensors cannot be known without accessing the UAV's internal systems, this paper demonstrates how an attacker can make the internal estimated value converge on the value that the attacker wants through spoofing, thereby indirectly getting the estimated values. In the attack model, it is assumed that as a minimum condition for spoofing attack, even if the attacker cannot access the internal of the UAV, the attacker can measure the true position of the UAV from the outside.

\begin{algorithm}
\caption{Proposed GPS Spoofing Attack Strategy}\label{alg:gps_spoofing}
\begin{algorithmic}
\STATE $UAV Mission$
\STATE $n \gets 0$
\WHILE{$!done$}
\STATE $attacker \, sets \, \Delta p_{n}$
\FOR{$t=0$ to $T$}
\STATE $sig_{n} \gets pos_{uav} + \sum_{i=0}^{n} \Delta p_{i}$
\STATE $transmit \, sig_{n}$
\ENDFOR
\STATE $n \gets n + 1$
\ENDWHILE
\end{algorithmic}
\end{algorithm}

Algorithm \ref{alg:gps_spoofing} presents a methodology for conducting a spoofing attack by utilizing the convergence of estimated values. $n$ is the step unit in which the DRL system determines the control action. $\Delta p_{n}$ represents the degree of change in the UAV's estimated position that the attacker desires for one step. $T$ is the number of spoofing attacks required to reflect this change in UAV. $\sum_{i=0}^{n} \Delta p_{i}$ is the total spoofed value reflected in the internal estimates of PX4 over n steps. $pos_{uav}$ is the current position of the UAV. In other words, $T$ spoofing attacks are required for $pos_{uav}$ to be reflected, and afterwards, the attacker must continuously add these reflected changes to the UAV's current position. The condition of the attacker model for a spoofing attack without detection by UAV is that the attacker can observe the actual position of the UAV from the outside. Therefore, $pos_{obs}$ can be considered almost identical to the UAV's actual position. In the experiments of this paper, it was demonstrated that approximately 30 spoofing attacks can change the latitude and longitude by 0.00003 each. Generally, in UAVs, the process of receiving and updating GPS signals occurs about 10 times per second which is the default setting for PX4. In conclusion, since a change of 0.00001 degrees in latitude or longitude is approximately equal to 1 meter, this means that it is possible to change the relative position inputs of DRL system by up to 1 meter along the x and y axes every 1 seconds.

This paper proposes a methodology for inducing collisions in a practical attack model. The practical attack model indicates that the internal estimated position of PX4 can be changed by 0.1 meter every 0.1 seconds. The internal estimated position determines the relative position, the input value for the DRL system in Equation \ref{eq:realtive}. Due to the narrower attack range compared to unconstrained attack model, this paper has induced actions that continuously move towards the collision object. Generally, if within a certain distance from the target, the optimized DRL system will take actions that reflect the momentum needed to move towards the target according to the hypothesis of this paper Section \ref{subsec:principle}. This paper proposes that when the input values of relative position are within a certain range and the UAV is near the collision object, the attacker uses vector values that align with the direction of the collision object to conduct spoofing attacks on the relative position and direction inputs. In other words, the attacker continuously injects spoofing attack vectors into the UAV, creating an effect similar to that of an obstacle exerting an attractive potential field on the UAV.

\begin{algorithm}
\caption{Constrained Attack}\label{alg:practical}
\begin{algorithmic}
\STATE $UAV Mission$
\STATE $n \gets 0$
\WHILE{$!done$}
\STATE $\Delta p_{n} \gets norm(pos_{col} - pos_{tar})$
\FOR{$t=0$ to $T$}
\STATE $sig_{n} \gets pos_{uav} + \sum_{i=0}^{n} \Delta p_{i}$
\STATE $transmit \, sig_{n}$
\ENDFOR
\STATE $pos_{est} \to sig_{n} \, (convergence)$
\STATE $UAV action$
\STATE $n \gets n + 1$
\ENDWHILE
\end{algorithmic}
\end{algorithm}

Algorithm \ref{alg:practical} represents the process of the proposed spoofing methodology which functions equivalently to the effect of the potential field created by obstacles. In this methodology, it is assumed that DRL outputs control actions at intervals of 0.1 seconds. The attacker performs spoofing T times to change the UAV's estimated value by $\Delta p$ every 0.1 seconds. Applying a change of vector $norm(pos_{col} - pos_{tar})$ to $pos_{est}$ ultimately causes relative position and direction ($d_{rel}, yaw_{rel}$) to the target to change in the direction towards the obstacle. The positions of the collision obstacle and the target are represented by $pos_{col}$ and $pos_{tar}$, respectively. $\pi$ is the DRL navigation system which outputs the control action. In our proposed attack method, uniform field $norm(pos_{col} - pos_{tar})$ are applied to spoofing attack for inducing the UAV to obstacle.

\section{Experiment}
\begin{figure}[]
\centering 
\begin{subfigure}{\linewidth}
\centering
\includegraphics[width=7.3cm, height=3.6cm]{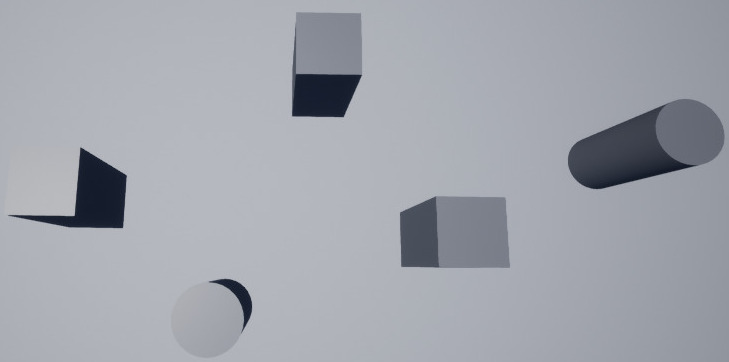}
\caption{Top view of flight environment map used in the experiment.\label{fig:env}}
 \end{subfigure}
 \vfill
\begin{subfigure}{\linewidth}
\centering
\includegraphics[width=5.5cm, height=5.5cm]{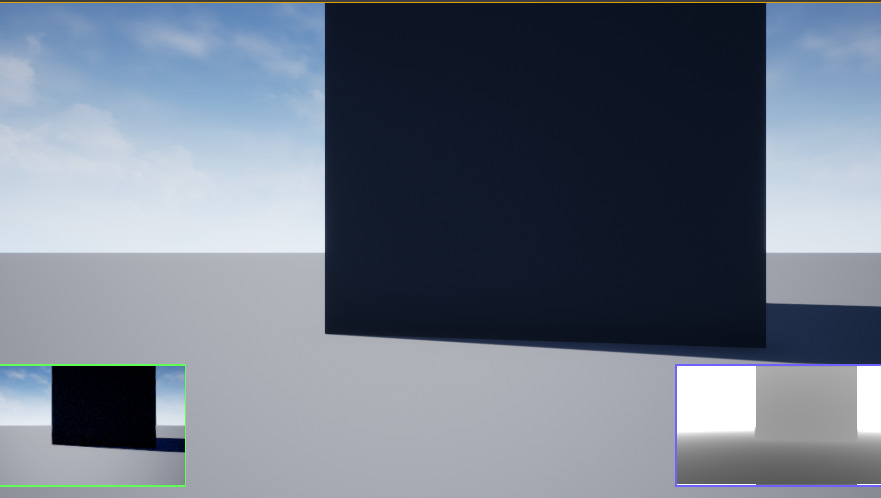}
\caption{Front view of flight environment map used in the experiment.\label{fig:front}}
 \end{subfigure}
  \caption{Flight experiment environment in the AirSim simulator.}
\end{figure}

AirSim is a highly realistic flight simulator based on Unreal Engine which provides physically and visually realistic environments \cite{airsim2017fsr}. In particular, it is optimized to provide an environment for AI research such as deep learning, computer vision, and reinforcement learning for UAVs. AirSim provides photo-realistic environments and delivers real-time simulation data through a Python API, enabling AI system to effectively communicate with the UAV during each simulation frame. This integration allows researchers to develop and test their algorithms in a highly realistic setting. The target of the attack in this paper is a DRL-based navigation system which is trained using TD3 based on the structures in \cite{he2020deep, he2021explainable}. DRL training and an agent is primarily based on the stable baseline library \cite{stable-baselines3}. The DRL system basically receives as input a depth image, the target's position, the current UAV's position, and current yaw. The DRL system outputs the speed and the angle of yaw to be changed based on current direction. These are dynamically obtained at each step through AirSim's Python API. The DRL system receives input every 0.1 seconds and then proceeds with the control action step. Figures \ref{fig:env} and \ref{fig:front} represent the flight environment from the top view and front view, respectively.

Lastly, to demonstrate the range of spoofing attack possible in the EKF system of an UAV autopilot system, PX4 and Gazebo simulation system were utilized \cite{px4, gazebo}.

\subsection{Unconstrained Attack Model}

\begin{figure}
\centering
\includegraphics[width=7.5cm, height=4.5cm]{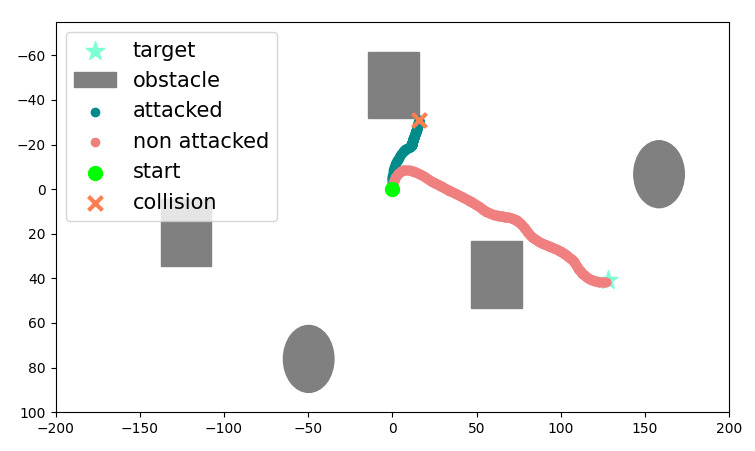}
\caption{This figure presents the results of inducing a collision using unconstrained attack methodology. The green line represents the path when the attack is carried out, while the pink line shows the path without the attack.\label{fig:theo}}
\end{figure}

In this experiment, we demonstrate that a spoofing attack can induce a collision using the methodology described in subsection \ref{sub:General}. The assumption of this experiment is that there are no restrictions on the inputs of the DRL system. Therefore, for all input spaces, an attacker can choose the desired output value, which leads to inducing collisions. Figure \ref{fig:theo} shows the experimental results. Both the green line and the purple line represent the paths of UAVs that were given the mission to reach the target. However, the green line shows the path when a spoofing attack was carried out, while the purple line shows the path when no attack was conducted. When the spoofing attack was carried out, it was possible to induce a collision. When no attack was conducted, we can see that the UAV moved accurately to the target. The experimental results show that the attacker can completely reverse the UAV's original path and cause collision.

\subsection{GPS Spoofing Strategy in PX4}\label{sub:exp_spoofing}
\begin{figure}[H] 
\centering
\includegraphics[width=6.9cm, height=6.3cm]{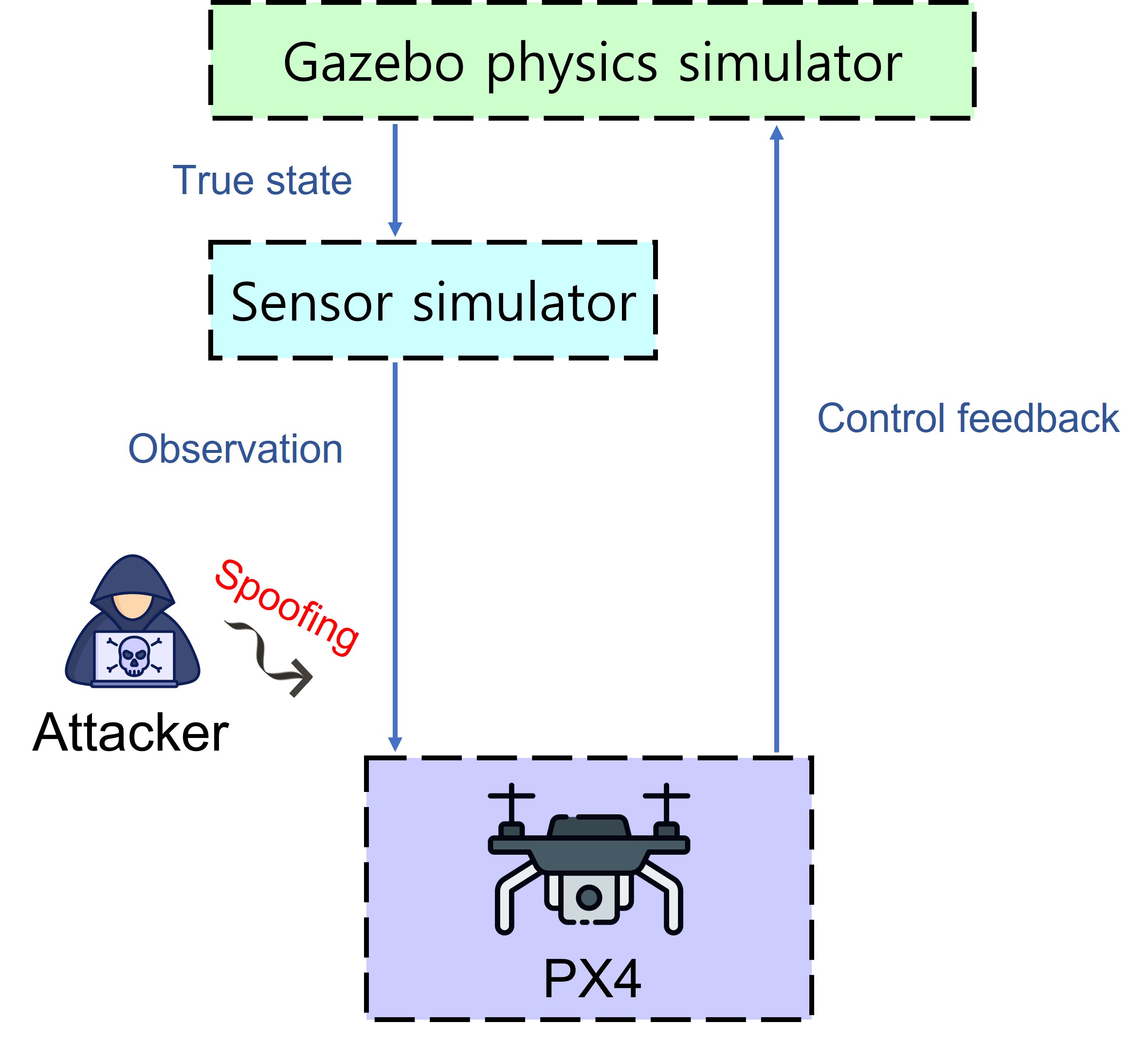}
\caption{The true values transmitted from the physical simulator are passed to sensor simulator and PX4 receives the simulated sensor value. To transmit spoofing values to PX4, this paper use the communication layer between sensor simulation and PX4.\label{fig:gazebo}}
\end{figure}

The objective of this experiment is to determine to what extent the spoofing attack model proposed in this paper can manipulate the UAV's estimated position per unit time. As mentioned earlier, in this experiment, we conducted tests on GPS spoofing, which is the most realistic and systematized spoofing attack method. The structure of the experiment involves examining how the internal estimates of PX4 are affected when spoofed values different from the actual position are input into PX4 autopilot system. To set up the environment where PX4 runs, we utilized Gazebo simulator. The overall experimental structure is shown in Figure \ref{fig:gazebo}.

\begin{figure}[H] 
\centering
\includegraphics[width=7cm, height=5.9cm]{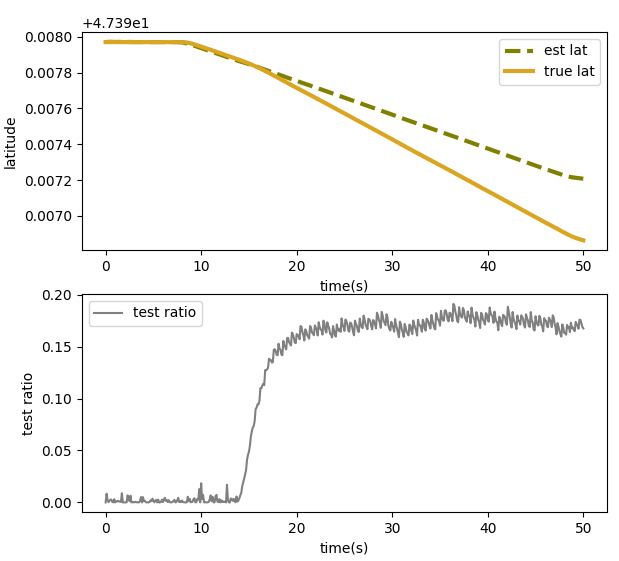}
\caption{This figure shows the results when a spoofing attack is conducted using the method proposed in this subsection. The orange line represents the actual position of the UAV, while the blue line indicates the position estimated internally by the UAV. The figure below shows the change in the test ratio value during the spoofing attack, and it can be observed that it does not exceed 1.\label{fig:lat}}
\end{figure}

The estimation of EKF sensor fusion determines the anomaly of sensor values from external sensor sources by comparing them with the current internal estimated values, and then reflects these values. At the initial stage of the attack, the internal estimated value will have little difference from the actually observed position, but as the attack progresses, it will diverge from the externally observed position. Therefore, the attacker must continuously predict the internal estimated value to be able to perform spoofing. Therefore, the attacker must be able to predict how the internal estimated value changes when the spoofing value is applied. To achieve this, the attacker needs to find a value that causes the change in the internal estimated value to converge as the spoofing attack progresses. For this purpose, in this paper, we heuristically found the value to which the internal position converges through the external position and spoofing attack values. The basic principle is referenced in \cite{jung2024analysis}. In this experiment, for every 30 GPS samples, the attacker changed both latitude and longitude by 0.0003 in the desired direction. This represents the time it takes for the change applied to the internal estimated value over 30 samples to converge. Figure \ref{fig:lat} shows the true position and the UAV's internal estimated position when the spoofing attack is executed starting from 15 seconds. A difference of approximately 0.00035 occurred between the true value and the estimated value over the period from 15 seconds to 50 seconds. This indicates that the spoofing values were continuously reflected, and additionally, during the same period, the test ratio maintained a value below 1 as shown in Figure \ref{fig:lat}. The results of this experiment demonstrated that it's possible to change the latitude by 0.0003 every 3 seconds. Simplifying this slightly, it means that a change of 0.1 meters can be made every 0.1 seconds because 1 degree of latitude or longitude equals to approximately 111 km.

\begin{figure}[]
\centering 
\begin{subfigure}{\linewidth}
\centering
\includegraphics[width=8cm, height=5.2cm]{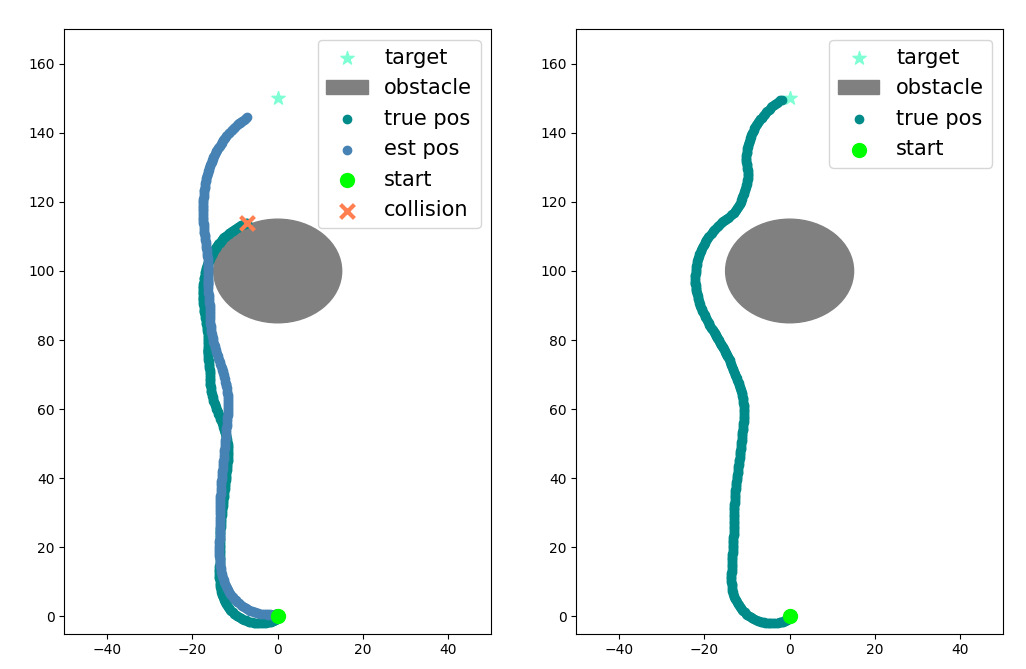}
\caption{Experimental results when the target (0, 150) is aligned with the obstacle (0, 100).}
 \label{sub:a}
 \end{subfigure}
 \vfill
\begin{subfigure}{\linewidth}
\centering
 \includegraphics[width=8cm, height=4.8cm]{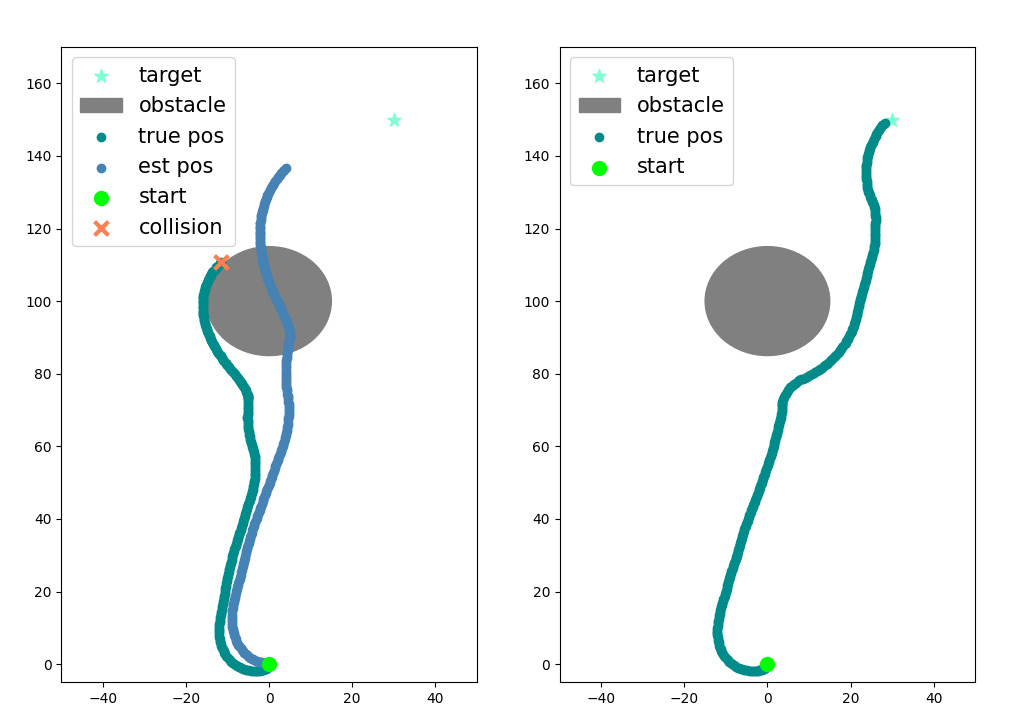}
 \caption{Experimental results when the target (30, 150) is positioned to the side of the obstacle (0, 100).}
 \end{subfigure}
   \caption{This experiment demonstrates the results of a spoofing attack method using a 0.1 m size vector to spoof the position sensor. In each figure, the left side shows the path when the spoofing attack is applied, while the right side shows the result without the spoofing attack. In the left figure, the blue line represents the internally estimated position that the UAV incorrectly recognizes due to spoofing, which differs from its actual position.}\label{fig:exp_practical}
\end{figure}

\subsection{Constrained Attack Model}

Based on the experimental results in subsection \ref{sub:exp_spoofing}, This paper presents a GPS spoofing attack model that closely simulates real-world scenarios. Specifically, given that the DRL model in this experiment operates every 0.1 seconds, the spoofing attack model is designed to assume a possible attack range at intervals of 0.1 seconds. In conclusion, the objective of this experiment was to demonstrate an attacker could induce a collision by assuming that, at each step, the attacker can apply a change of 0.1 meters in the form of a vector that the DRL system recognizes. The experimental environment assumes that the UAV is positioned 100 meters away from the obstacle, with the target located at a distance of 150 meters, allowing for a flight simulation based on this mission. At this point, the attack methodology employed the method used in subsection \ref{sub:practical}. Figure \ref{fig:exp_practical} shows the results of the experiment. The upper part shows the results when the target is positioned at (0, 150 m). the left side shows the results when an attack was conducted, while the right side shows the results when no attack occurred. The lower part shows the results when the target is positioned at (30 m, 150 m). The green path indicates the true position of movement, while the blue path in the left represents the position recognized by the UAV. The experimental results show that the attacker can induce a collision not only when the UAV, target, and obstacle are aligned in a straight line, but also when the target is positioned to the side of the obstacle.

\section{Discussion}

The most common assumption in AI system attacks involves methods that don't separately consider lower-layer components. These approaches focus solely on the inputs and control outputs of DRL-based navigation system without constraints, allowing attackers to freely manipulate positional or directional values. This inherently makes such attacks easier to execute, which experimental results naturally reflect. However, no prior study has yet demonstrated collision induction through positional spoofing attacks in DRL-based navigation systems for UAV. This paper's significance lies in proposing such an attack model.

To overcome the limitations of unrestricted attacks discussed earlier, this paper first models the feasible range of GPS spoofing at the UAV navigation layer. subsection \ref{sub:exp_spoofing} presents these findings: When the test ratio exceeds 1, PX4’s state estimation module flags it as an outlier. However, GPS spoofing can evade detection by maintaining values below 0.2. Note that such spoofing has inherent motion constraints for UAVs, with spoofable displacements limited to 0.1m per 0.1s. Under these constraints, subsection \ref{Practical Attack Model} experimentally validates how attackers can still maximize collision induction efficiency within restricted DRL parameters. Crucially, our method generates a potential field based on the relative direction between the collision target and the UAV’s position during GPS spoofing, proving sufficient for reliable collision guidance. Nevertheless, this study has certain limitations. Due to experimental difficulties in directly integrating DRL systems with GPS spoofing, this paper adopted an alternative approach of modeling each attack separately while imposing constraints on the DRL system's inputs. However, this paper contributes to more realistic attacker modeling at both the AI system's module layer and the control system's state estimation layer compared to previous work. Our approach provides a practical attack scenario that could benefit future AI system security research. A fundamental limitation in practical GPS spoofing implementation remains the prerequisite requirement of knowing the UAV's precise position through radar or visual observation. Despite these constraints, our assumptions carry significant implications for future research, given the current lack of practical methods to bypass anomaly detection in state estimation systems.

\section{Conclusion}

There is a growing trend of research on the AI-based UAV navigation system. However, there is very little research on the security aspects of such systems. Despite recent important security papers focusing on sensor attacks using adversarial methods \cite{hickling2023robust}, there is still very little research on the security aspects of these systems. Adversarial attacks still have limitations in leading to practical attacks, such as physically injecting subtle noise into sensors. This paper proposes a systematic and more practical attack model based on GPS spoofing attack. Most studies on GPS spoofing attacks against UAVs have assumed attacks on controllers with simple missions, such as moving to a specific location. These studies have focused on changing the UAV's movement or inducing failsafe mechanisms \cite{sathaye2022semperfi, agyapong2021efficient, jullian2021deep, noh2019tractor, sathaye2022experimental}. This paper proposes GPS spoofing attack model for intelligent UAVs equipped with obstacle avoidance capabilities. To achieve this, this paper analyzed structural issues in the input space of state-of-the-art the DRL-based systems and experimentally demonstrated that attacks are possible through only spoofing position value. Unrestricted attacker model has conditions that are just as stringent as those in traditional adversarial attacks. The attacker must know not only the parameters of the DRL policy but also the current position of the UAV and the depth image of the direction the UAV is facing. While this is a theoretically possible attack scenario, it becomes a challenging scenario in practice. Unlike unconstrained model, the attacker of constrained model only needs to know the position of the UAV. This presents a new attacker model by combining sensor spoofing attacks in DRL system with the limitations of EKF module in PX4 autopilot system. This proposes a more realistic attack threat and becomes a crucial point in future DRL-based navigation system designs. Subsection \ref{subsec:principle} analyzes the vulnerabilities of the most efficient DRL navigation model in recent research. Furthermore, the experiments in this paper conducted attacks on the recently researched DRL systems, thereby validating the attack model proposed in this study. However, this paper has only demonstrated the success of the attack model using uniform vectors, and has not yet presented the efficiency of the methodology using various other potential vectors. These present promising research topics for future work.

%% BioMed_Central_Bib_Style_v1.01


\begin{thebibliography}{44}
% BibTex style file: bmc-mathphys.bst (version 2.1), 2014-07-24
\ifx \bisbn   \undefined \def \bisbn  #1{ISBN #1}\fi
\ifx \binits  \undefined \def \binits#1{#1}\fi
\ifx \bauthor  \undefined \def \bauthor#1{#1}\fi
\ifx \batitle  \undefined \def \batitle#1{#1}\fi
\ifx \bjtitle  \undefined \def \bjtitle#1{#1}\fi
\ifx \bvolume  \undefined \def \bvolume#1{\textbf{#1}}\fi
\ifx \byear  \undefined \def \byear#1{#1}\fi
\ifx \bissue  \undefined \def \bissue#1{#1}\fi
\ifx \bfpage  \undefined \def \bfpage#1{#1}\fi
\ifx \blpage  \undefined \def \blpage #1{#1}\fi
\ifx \burl  \undefined \def \burl#1{\textsf{#1}}\fi
\ifx \doiurl  \undefined \def \doiurl#1{\url{https://doi.org/#1}}\fi
\ifx \betal  \undefined \def \betal{\textit{et al.}}\fi
\ifx \binstitute  \undefined \def \binstitute#1{#1}\fi
\ifx \binstitutionaled  \undefined \def \binstitutionaled#1{#1}\fi
\ifx \bctitle  \undefined \def \bctitle#1{#1}\fi
\ifx \beditor  \undefined \def \beditor#1{#1}\fi
\ifx \bpublisher  \undefined \def \bpublisher#1{#1}\fi
\ifx \bbtitle  \undefined \def \bbtitle#1{#1}\fi
\ifx \bedition  \undefined \def \bedition#1{#1}\fi
\ifx \bseriesno  \undefined \def \bseriesno#1{#1}\fi
\ifx \blocation  \undefined \def \blocation#1{#1}\fi
\ifx \bsertitle  \undefined \def \bsertitle#1{#1}\fi
\ifx \bsnm \undefined \def \bsnm#1{#1}\fi
\ifx \bsuffix \undefined \def \bsuffix#1{#1}\fi
\ifx \bparticle \undefined \def \bparticle#1{#1}\fi
\ifx \barticle \undefined \def \barticle#1{#1}\fi
\bibcommenthead
\ifx \bconfdate \undefined \def \bconfdate #1{#1}\fi
\ifx \botherref \undefined \def \botherref #1{#1}\fi
\ifx \url \undefined \def \url#1{\textsf{#1}}\fi
\ifx \bchapter \undefined \def \bchapter#1{#1}\fi
\ifx \bbook \undefined \def \bbook#1{#1}\fi
\ifx \bcomment \undefined \def \bcomment#1{#1}\fi
\ifx \oauthor \undefined \def \oauthor#1{#1}\fi
\ifx \citeauthoryear \undefined \def \citeauthoryear#1{#1}\fi
\ifx \endbibitem  \undefined \def \endbibitem {}\fi
\ifx \bconflocation  \undefined \def \bconflocation#1{#1}\fi
\ifx \arxivurl  \undefined \def \arxivurl#1{\textsf{#1}}\fi
\csname PreBibitemsHook\endcsname

%%% 1
\bibitem[\protect\citeauthoryear{Chmaj and Selvaraj}{2015}]{chmaj2015distributed}
\begin{bchapter}
\bauthor{\bsnm{Chmaj}, \binits{G.}},
\bauthor{\bsnm{Selvaraj}, \binits{H.}}:
\bctitle{Distributed processing applications for uav/drones: a survey}.
In: \bbtitle{Progress in Systems Engineering: Proceedings of the Twenty-Third International Conference on Systems Engineering},
pp. \bfpage{449}--\blpage{454}
(\byear{2015}).
\bcomment{Springer}
\end{bchapter}
\endbibitem

%%% 2
\bibitem[\protect\citeauthoryear{Restas}{2015}]{restas2015drone}
\begin{barticle}
\bauthor{\bsnm{Restas}, \binits{A.}}:
\batitle{Drone applications for supporting disaster management}.
\bjtitle{World Journal of Engineering and Technology}
\bvolume{3}(\bissue{3}),
\bfpage{316}--\blpage{321}
(\byear{2015})
\end{barticle}
\endbibitem

%%% 3
\bibitem[\protect\citeauthoryear{Busby}{2019}]{busby2019drone}
\begin{barticle}
\bauthor{\bsnm{Busby}, \binits{J.}}:
\batitle{Drone delivery: The danger of opening the air as a commercial highway}.
\bjtitle{Loy. Mar. LJ}
\bvolume{18},
\bfpage{287}
(\byear{2019})
\end{barticle}
\endbibitem

%%% 4
\bibitem[\protect\citeauthoryear{Jung and Kim}{2017}]{jung2017analysis}
\begin{barticle}
\bauthor{\bsnm{Jung}, \binits{S.}},
\bauthor{\bsnm{Kim}, \binits{H.}}:
\batitle{Analysis of amazon prime air uav delivery service}.
\bjtitle{Journal of Knowledge Information Technology and Systems}
\bvolume{12}(\bissue{2}),
\bfpage{253}--\blpage{266}
(\byear{2017})
\end{barticle}
\endbibitem

%%% 5
\bibitem[\protect\citeauthoryear{Paucar et~al.}{2018}]{paucar2018use}
\begin{bchapter}
\bauthor{\bsnm{Paucar}, \binits{C.}},
\bauthor{\bsnm{Morales}, \binits{L.}},
\bauthor{\bsnm{Pinto}, \binits{K.}},
\bauthor{\bsnm{S{\'a}nchez}, \binits{M.}},
\bauthor{\bsnm{Rodr{\'\i}guez}, \binits{R.}},
\bauthor{\bsnm{Gutierrez}, \binits{M.}},
\bauthor{\bsnm{Palacios}, \binits{L.}}:
\bctitle{Use of drones for surveillance and reconnaissance of military areas}.
In: \bbtitle{Developments and Advances in Defense and Security: Proceedings of the Multidisciplinary International Conference of Research Applied to Defense and Security (MICRADS 2018)},
pp. \bfpage{119}--\blpage{132}
(\byear{2018}).
\bcomment{Springer}
\end{bchapter}
\endbibitem

%%% 6
\bibitem[\protect\citeauthoryear{Luckey}{2022}]{luckey2022anduril}
\begin{barticle}
\bauthor{\bsnm{Luckey}, \binits{P.}}:
\batitle{Anduril founder and ceo palmer luckey in conversation with apdr editor kym bergmann}.
\bjtitle{Asia-Pacific Defence Reporter (2002)}
\bvolume{48}(\bissue{8}),
\bfpage{28}--\blpage{31}
(\byear{2022})
\end{barticle}
\endbibitem

%%% 7
\bibitem[\protect\citeauthoryear{Dew and Lewis}{2024}]{dew2024us}
\begin{barticle}
\bauthor{\bsnm{Dew}, \binits{N.}},
\bauthor{\bsnm{Lewis}, \binits{I.}}:
\batitle{Us defense innovation and industrial policy: An assessment of where things currently stand}.
\bjtitle{Expeditions with MCUP}
\bvolume{2024}(\bissue{1}),
\bfpage{1}--\blpage{25}
(\byear{2024})
\end{barticle}
\endbibitem

%%% 8
\bibitem[\protect\citeauthoryear{Chen et~al.}{2022}]{chen2022end}
\begin{barticle}
\bauthor{\bsnm{Chen}, \binits{S.}},
\bauthor{\bsnm{Zhou}, \binits{W.}},
\bauthor{\bsnm{Yang}, \binits{A.-S.}},
\bauthor{\bsnm{Chen}, \binits{H.}},
\bauthor{\bsnm{Li}, \binits{B.}},
\bauthor{\bsnm{Wen}, \binits{C.-Y.}}:
\batitle{An end-to-end uav simulation platform for visual slam and navigation}.
\bjtitle{Aerospace}
\bvolume{9}(\bissue{2}),
\bfpage{48}
(\byear{2022})
\end{barticle}
\endbibitem

%%% 9
\bibitem[\protect\citeauthoryear{Mac et~al.}{2016}]{mac2016improved}
\begin{bchapter}
\bauthor{\bsnm{Mac}, \binits{T.T.}},
\bauthor{\bsnm{Copot}, \binits{C.}},
\bauthor{\bsnm{Hernandez}, \binits{A.}},
\bauthor{\bsnm{De~Keyser}, \binits{R.}}:
\bctitle{Improved potential field method for unknown obstacle avoidance using uav in indoor environment}.
In: \bbtitle{2016 IEEE 14th International Symposium on Applied Machine Intelligence and Informatics (SAMI)},
pp. \bfpage{345}--\blpage{350}
(\byear{2016}).
\bcomment{IEEE}
\end{bchapter}
\endbibitem

%%% 10
\bibitem[\protect\citeauthoryear{Escobar-Alvarez et~al.}{2018}]{escobar2018r}
\begin{barticle}
\bauthor{\bsnm{Escobar-Alvarez}, \binits{H.D.}},
\bauthor{\bsnm{Johnson}, \binits{N.}},
\bauthor{\bsnm{Hebble}, \binits{T.}},
\bauthor{\bsnm{Klingebiel}, \binits{K.}},
\bauthor{\bsnm{Quintero}, \binits{S.A.}},
\bauthor{\bsnm{Regenstein}, \binits{J.}},
\bauthor{\bsnm{Browning}, \binits{N.A.}}:
\batitle{R-advance: rapid adaptive prediction for vision-based autonomous navigation, control, and evasion}.
\bjtitle{Journal of Field Robotics}
\bvolume{35}(\bissue{1}),
\bfpage{91}--\blpage{100}
(\byear{2018})
\end{barticle}
\endbibitem

%%% 11
\bibitem[\protect\citeauthoryear{Hickling et~al.}{2023}]{hickling2023robust}
\begin{botherref}
\oauthor{\bsnm{Hickling}, \binits{T.}},
\oauthor{\bsnm{Aouf}, \binits{N.}},
\oauthor{\bsnm{Spencer}, \binits{P.}}:
Robust adversarial attacks detection based on explainable deep reinforcement learning for uav guidance and planning.
IEEE Transactions on Intelligent Vehicles
(2023)
\end{botherref}
\endbibitem

%%% 12
\bibitem[\protect\citeauthoryear{Sato et~al.}{2021}]{sato2021dirty}
\begin{bchapter}
\bauthor{\bsnm{Sato}, \binits{T.}},
\bauthor{\bsnm{Shen}, \binits{J.}},
\bauthor{\bsnm{Wang}, \binits{N.}},
\bauthor{\bsnm{Jia}, \binits{Y.}},
\bauthor{\bsnm{Lin}, \binits{X.}},
\bauthor{\bsnm{Chen}, \binits{Q.A.}}:
\bctitle{Dirty road can attack: Security of deep learning based automated lane centering under $\{$Physical-World$\}$ attack}.
In: \bbtitle{30th USENIX Security Symposium (USENIX Security 21)},
pp. \bfpage{3309}--\blpage{3326}
(\byear{2021})
\end{bchapter}
\endbibitem

%%% 13
\bibitem[\protect\citeauthoryear{Zhong et~al.}{2022}]{zhong2022neural}
\begin{barticle}
\bauthor{\bsnm{Zhong}, \binits{Z.}},
\bauthor{\bsnm{Kaiser}, \binits{G.}},
\bauthor{\bsnm{Ray}, \binits{B.}}:
\batitle{Neural network guided evolutionary fuzzing for finding traffic violations of autonomous vehicles}.
\bjtitle{IEEE Transactions on Software Engineering}
\bvolume{49}(\bissue{4}),
\bfpage{1860}--\blpage{1875}
(\byear{2022})
\end{barticle}
\endbibitem

%%% 14
\bibitem[\protect\citeauthoryear{Zhou et~al.}{2022}]{zhou2022strategic}
\begin{bchapter}
\bauthor{\bsnm{Zhou}, \binits{X.}},
\bauthor{\bsnm{Schmedding}, \binits{A.}},
\bauthor{\bsnm{Ren}, \binits{H.}},
\bauthor{\bsnm{Yang}, \binits{L.}},
\bauthor{\bsnm{Schowitz}, \binits{P.}},
\bauthor{\bsnm{Smirni}, \binits{E.}},
\bauthor{\bsnm{Alemzadeh}, \binits{H.}}:
\bctitle{Strategic safety-critical attacks against an advanced driver assistance system}.
In: \bbtitle{2022 52nd Annual IEEE/IFIP International Conference on Dependable Systems and Networks (DSN)},
pp. \bfpage{79}--\blpage{87}
(\byear{2022}).
\bcomment{IEEE}
\end{bchapter}
\endbibitem

%%% 15
\bibitem[\protect\citeauthoryear{Quinonez et~al.}{2020}]{quinonez2020savior}
\begin{bchapter}
\bauthor{\bsnm{Quinonez}, \binits{R.}},
\bauthor{\bsnm{Giraldo}, \binits{J.}},
\bauthor{\bsnm{Salazar}, \binits{L.}},
\bauthor{\bsnm{Bauman}, \binits{E.}},
\bauthor{\bsnm{Cardenas}, \binits{A.}},
\bauthor{\bsnm{Lin}, \binits{Z.}}:
\bctitle{$\{$SAVIOR$\}$: Securing autonomous vehicles with robust physical invariants}.
In: \bbtitle{29th USENIX Security Symposium (USENIX Security 20)},
pp. \bfpage{895}--\blpage{912}
(\byear{2020})
\end{bchapter}
\endbibitem

%%% 16
\bibitem[\protect\citeauthoryear{Shen et~al.}{2020}]{shen2020drift}
\begin{bchapter}
\bauthor{\bsnm{Shen}, \binits{J.}},
\bauthor{\bsnm{Won}, \binits{J.Y.}},
\bauthor{\bsnm{Chen}, \binits{Z.}},
\bauthor{\bsnm{Chen}, \binits{Q.A.}}:
\bctitle{Drift with devil: Security of $\{$Multi-Sensor$\}$ fusion based localization in $\{$High-Level$\}$ autonomous driving under $\{$GPS$\}$ spoofing}.
In: \bbtitle{29th USENIX Security Symposium (USENIX Security 20)},
pp. \bfpage{931}--\blpage{948}
(\byear{2020})
\end{bchapter}
\endbibitem

%%% 17
\bibitem[\protect\citeauthoryear{Khan et~al.}{2024}]{khan2024comprehensive}
\begin{bchapter}
\bauthor{\bsnm{Khan}, \binits{A.}},
\bauthor{\bsnm{Ivaki}, \binits{N.}},
\bauthor{\bsnm{Madeira}, \binits{H.}}:
\bctitle{A comprehensive study on drones resilience in the presence of inertial measurement unit faults}.
In: \bbtitle{2024 54th Annual IEEE/IFIP International Conference on Dependable Systems and Networks (DSN)},
pp. \bfpage{316}--\blpage{323}
(\byear{2024}).
\bcomment{IEEE}
\end{bchapter}
\endbibitem

%%% 18
\bibitem[\protect\citeauthoryear{Li}{2017}]{li2017deep}
\begin{botherref}
\oauthor{\bsnm{Li}, \binits{Y.}}:
Deep reinforcement learning: An overview.
arXiv preprint arXiv:1701.07274
(2017)
\end{botherref}
\endbibitem

%%% 19
\bibitem[\protect\citeauthoryear{He et~al.}{2020}]{he2020deep}
\begin{botherref}
\oauthor{\bsnm{He}, \binits{L.}},
\oauthor{\bsnm{Aouf}, \binits{N.}},
\oauthor{\bsnm{Whidborne}, \binits{J.F.}},
\oauthor{\bsnm{Song}, \binits{B.}}:
Deep reinforcement learning based local planner for uav obstacle avoidance using demonstration data.
arXiv preprint arXiv:2008.02521
(2020)
\end{botherref}
\endbibitem

%%% 20
\bibitem[\protect\citeauthoryear{Lillicrap}{2015}]{lillicrap2015continuous}
\begin{botherref}
\oauthor{\bsnm{Lillicrap}, \binits{T.}}:
Continuous control with deep reinforcement learning.
arXiv preprint arXiv:1509.02971
(2015)
\end{botherref}
\endbibitem

%%% 21
\bibitem[\protect\citeauthoryear{Mnih}{2013}]{mnih2013playing}
\begin{botherref}
\oauthor{\bsnm{Mnih}, \binits{V.}}:
Playing atari with deep reinforcement learning.
arXiv preprint arXiv:1312.5602
(2013)
\end{botherref}
\endbibitem

%%% 22
\bibitem[\protect\citeauthoryear{Fujimoto et~al.}{2018}]{fujimoto2018addressing}
\begin{bchapter}
\bauthor{\bsnm{Fujimoto}, \binits{S.}},
\bauthor{\bsnm{Hoof}, \binits{H.}},
\bauthor{\bsnm{Meger}, \binits{D.}}:
\bctitle{Addressing function approximation error in actor-critic methods}.
In: \bbtitle{International Conference on Machine Learning},
pp. \bfpage{1587}--\blpage{1596}
(\byear{2018}).
\bcomment{PMLR}
\end{bchapter}
\endbibitem

%%% 23
\bibitem[\protect\citeauthoryear{Kalidas et~al.}{2023}]{kalidas2023deep}
\begin{barticle}
\bauthor{\bsnm{Kalidas}, \binits{A.P.}},
\bauthor{\bsnm{Joshua}, \binits{C.J.}},
\bauthor{\bsnm{Md}, \binits{A.Q.}},
\bauthor{\bsnm{Basheer}, \binits{S.}},
\bauthor{\bsnm{Mohan}, \binits{S.}},
\bauthor{\bsnm{Sakri}, \binits{S.}}:
\batitle{Deep reinforcement learning for vision-based navigation of uavs in avoiding stationary and mobile obstacles}.
\bjtitle{Drones}
\bvolume{7}(\bissue{4}),
\bfpage{245}
(\byear{2023})
\end{barticle}
\endbibitem

%%% 24
\bibitem[\protect\citeauthoryear{He et~al.}{2021}]{he2021explainable}
\begin{barticle}
\bauthor{\bsnm{He}, \binits{L.}},
\bauthor{\bsnm{Aouf}, \binits{N.}},
\bauthor{\bsnm{Song}, \binits{B.}}:
\batitle{Explainable deep reinforcement learning for uav autonomous path planning}.
\bjtitle{Aerospace science and technology}
\bvolume{118},
\bfpage{107052}
(\byear{2021})
\end{barticle}
\endbibitem

%%% 25
\bibitem[\protect\citeauthoryear{Krishnan et~al.}{2021}]{krishnan2021air}
\begin{barticle}
\bauthor{\bsnm{Krishnan}, \binits{S.}},
\bauthor{\bsnm{Boroujerdian}, \binits{B.}},
\bauthor{\bsnm{Fu}, \binits{W.}},
\bauthor{\bsnm{Faust}, \binits{A.}},
\bauthor{\bsnm{Reddi}, \binits{V.J.}}:
\batitle{Air learning: a deep reinforcement learning gym for autonomous aerial robot visual navigation}.
\bjtitle{Machine Learning}
\bvolume{110}(\bissue{9}),
\bfpage{2501}--\blpage{2540}
(\byear{2021})
\end{barticle}
\endbibitem

%%% 26
\bibitem[\protect\citeauthoryear{Cetin et~al.}{2019}]{cetin2019drone}
\begin{bchapter}
\bauthor{\bsnm{Cetin}, \binits{E.}},
\bauthor{\bsnm{Barrado}, \binits{C.}},
\bauthor{\bsnm{Mu{\~n}oz}, \binits{G.}},
\bauthor{\bsnm{Macias}, \binits{M.}},
\bauthor{\bsnm{Pastor}, \binits{E.}}:
\bctitle{Drone navigation and avoidance of obstacles through deep reinforcement learning}.
In: \bbtitle{2019 IEEE/AIAA 38th Digital Avionics Systems Conference (DASC)},
pp. \bfpage{1}--\blpage{7}
(\byear{2019}).
\bcomment{IEEE}
\end{bchapter}
\endbibitem

%%% 27
\bibitem[\protect\citeauthoryear{Zhang et~al.}{2022}]{zhang2022autonomous}
\begin{barticle}
\bauthor{\bsnm{Zhang}, \binits{S.}},
\bauthor{\bsnm{Li}, \binits{Y.}},
\bauthor{\bsnm{Dong}, \binits{Q.}}:
\batitle{Autonomous navigation of uav in multi-obstacle environments based on a deep reinforcement learning approach}.
\bjtitle{Applied Soft Computing}
\bvolume{115},
\bfpage{108194}
(\byear{2022})
\end{barticle}
\endbibitem

%%% 28
\bibitem[\protect\citeauthoryear{Kalman}{1960}]{kalman1960new}
\begin{botherref}
\oauthor{\bsnm{Kalman}, \binits{R.E.}}:
A new approach to linear filtering and prediction problems
(1960)
\end{botherref}
\endbibitem

%%% 29
\bibitem[\protect\citeauthoryear{Team}{2024}]{px4}
\begin{botherref}
\oauthor{\bsnm{Team}, \binits{P.D.}}:
PX4 Autopilot.
\url{https://docs.px4.io}.
Accessed: 2024-10-01
(2024)
\end{botherref}
\endbibitem

%%% 30
\bibitem[\protect\citeauthoryear{Liu et~al.}{2019}]{liu2019analysis}
\begin{barticle}
\bauthor{\bsnm{Liu}, \binits{Y.}},
\bauthor{\bsnm{Li}, \binits{S.}},
\bauthor{\bsnm{Fu}, \binits{Q.}},
\bauthor{\bsnm{Liu}, \binits{Z.}},
\bauthor{\bsnm{Zhou}, \binits{Q.}}:
\batitle{Analysis of kalman filter innovation-based gnss spoofing detection method for ins/gnss integrated navigation system}.
\bjtitle{IEEE Sensors Journal}
\bvolume{19}(\bissue{13}),
\bfpage{5167}--\blpage{5178}
(\byear{2019})
\end{barticle}
\endbibitem

%%% 31
\bibitem[\protect\citeauthoryear{Noh et~al.}{2019}]{noh2019tractor}
\begin{barticle}
\bauthor{\bsnm{Noh}, \binits{J.}},
\bauthor{\bsnm{Kwon}, \binits{Y.}},
\bauthor{\bsnm{Son}, \binits{Y.}},
\bauthor{\bsnm{Shin}, \binits{H.}},
\bauthor{\bsnm{Kim}, \binits{D.}},
\bauthor{\bsnm{Choi}, \binits{J.}},
\bauthor{\bsnm{Kim}, \binits{Y.}}:
\batitle{Tractor beam: Safe-hijacking of consumer drones with adaptive gps spoofing}.
\bjtitle{ACM Transactions on Privacy and Security (TOPS)}
\bvolume{22}(\bissue{2}),
\bfpage{1}--\blpage{26}
(\byear{2019})
\end{barticle}
\endbibitem

%%% 32
\bibitem[\protect\citeauthoryear{Jung et~al.}{2024}]{jung2024analysis}
\begin{botherref}
\oauthor{\bsnm{Jung}, \binits{J.H.}},
\oauthor{\bsnm{Hong}, \binits{M.Y.}},
\oauthor{\bsnm{Choi}, \binits{H.}},
\oauthor{\bsnm{Yoon}, \binits{J.W.}}:
An analysis of gps spoofing attack and efficient approach to spoofing detection in px4.
IEEE Access
(2024)
\end{botherref}
\endbibitem

%%% 33
\bibitem[\protect\citeauthoryear{Feng et~al.}{2018}]{feng2018efficient}
\begin{barticle}
\bauthor{\bsnm{Feng}, \binits{Z.}},
\bauthor{\bsnm{Guan}, \binits{N.}},
\bauthor{\bsnm{Lv}, \binits{M.}},
\bauthor{\bsnm{Liu}, \binits{W.}},
\bauthor{\bsnm{Deng}, \binits{Q.}},
\bauthor{\bsnm{Liu}, \binits{X.}},
\bauthor{\bsnm{Yi}, \binits{W.}}:
\batitle{An efficient uav hijacking detection method using onboard inertial measurement unit}.
\bjtitle{ACM Transactions on Embedded Computing Systems (TECS)}
\bvolume{17}(\bissue{6}),
\bfpage{1}--\blpage{19}
(\byear{2018})
\end{barticle}
\endbibitem

%%% 34
\bibitem[\protect\citeauthoryear{Tippenhauer et~al.}{2011}]{tippenhauer2011requirements}
\begin{bchapter}
\bauthor{\bsnm{Tippenhauer}, \binits{N.O.}},
\bauthor{\bsnm{P{\"o}pper}, \binits{C.}},
\bauthor{\bsnm{Rasmussen}, \binits{K.B.}},
\bauthor{\bsnm{Capkun}, \binits{S.}}:
\bctitle{On the requirements for successful gps spoofing attacks}.
In: \bbtitle{Proceedings of the 18th ACM Conference on Computer and Communications Security},
pp. \bfpage{75}--\blpage{86}
(\byear{2011})
\end{bchapter}
\endbibitem

%%% 35
\bibitem[\protect\citeauthoryear{Sathaye et~al.}{2022}]{sathaye2022experimental}
\begin{bchapter}
\bauthor{\bsnm{Sathaye}, \binits{H.}},
\bauthor{\bsnm{Strohmeier}, \binits{M.}},
\bauthor{\bsnm{Lenders}, \binits{V.}},
\bauthor{\bsnm{Ranganathan}, \binits{A.}}:
\bctitle{An experimental study of $\{$GPS$\}$ spoofing and takeover attacks on $\{$UAVs$\}$}.
In: \bbtitle{31st USENIX Security Symposium (USENIX Security 22)},
pp. \bfpage{3503}--\blpage{3520}
(\byear{2022})
\end{bchapter}
\endbibitem

%%% 36
\bibitem[\protect\citeauthoryear{Chen et~al.}{2023}]{chen2023magnetic}
\begin{barticle}
\bauthor{\bsnm{Chen}, \binits{B.}},
\bauthor{\bsnm{Huang}, \binits{L.}},
\bauthor{\bsnm{Zhang}, \binits{K.}},
\bauthor{\bsnm{Hu}, \binits{J.}},
\bauthor{\bsnm{Zhu}, \binits{W.}}:
\batitle{Magnetic interference analysis and compensation method of airborne electronic equipment in an unmanned aerial vehicle}.
\bjtitle{Applied Sciences}
\bvolume{13}(\bissue{13}),
\bfpage{7455}
(\byear{2023})
\end{barticle}
\endbibitem

%%% 37
\bibitem[\protect\citeauthoryear{Kim et~al.}{2024}]{kim2024systematic}
\begin{bchapter}
\bauthor{\bsnm{Kim}, \binits{H.}},
\bauthor{\bsnm{Bandyopadhyay}, \binits{R.}},
\bauthor{\bsnm{Ozmen}, \binits{M.O.}},
\bauthor{\bsnm{Celik}, \binits{Z.B.}},
\bauthor{\bsnm{Bianchi}, \binits{A.}},
\bauthor{\bsnm{Kim}, \binits{Y.}},
\bauthor{\bsnm{Xu}, \binits{D.}}:
\bctitle{A systematic study of physical sensor attack hardness}.
In: \bbtitle{2024 IEEE Symposium on Security and Privacy (SP)},
pp. \bfpage{143}--\blpage{143}
(\byear{2024}).
\bcomment{IEEE Computer Society}
\end{bchapter}
\endbibitem

%%% 38
\bibitem[\protect\citeauthoryear{Jang et~al.}{2023}]{jang2023paralyzing}
\begin{bchapter}
\bauthor{\bsnm{Jang}, \binits{J.-H.}},
\bauthor{\bsnm{Cho}, \binits{M.}},
\bauthor{\bsnm{Kim}, \binits{J.}},
\bauthor{\bsnm{Kim}, \binits{D.}},
\bauthor{\bsnm{Kim}, \binits{Y.}}:
\bctitle{Paralyzing drones via emi signal injection on sensory communication channels.}
In: \bbtitle{NDSS}
(\byear{2023})
\end{bchapter}
\endbibitem

%%% 39
\bibitem[\protect\citeauthoryear{Shah et~al.}{2017}]{airsim2017fsr}
\begin{bchapter}
\bauthor{\bsnm{Shah}, \binits{S.}},
\bauthor{\bsnm{Dey}, \binits{D.}},
\bauthor{\bsnm{Lovett}, \binits{C.}},
\bauthor{\bsnm{Kapoor}, \binits{A.}}:
\bctitle{Airsim: High-fidelity visual and physical simulation for autonomous vehicles}.
In: \bbtitle{Field and Service Robotics}
(\byear{2017}).
\burl{https://arxiv.org/abs/1705.05065}
\end{bchapter}
\endbibitem

%%% 40
\bibitem[\protect\citeauthoryear{Raffin et~al.}{2021}]{stable-baselines3}
\begin{barticle}
\bauthor{\bsnm{Raffin}, \binits{A.}},
\bauthor{\bsnm{Hill}, \binits{A.}},
\bauthor{\bsnm{Gleave}, \binits{A.}},
\bauthor{\bsnm{Kanervisto}, \binits{A.}},
\bauthor{\bsnm{Ernestus}, \binits{M.}},
\bauthor{\bsnm{Dormann}, \binits{N.}}:
\batitle{Stable-baselines3: Reliable reinforcement learning implementations}.
\bjtitle{Journal of Machine Learning Research}
\bvolume{22}(\bissue{268}),
\bfpage{1}--\blpage{8}
(\byear{2021})
\end{barticle}
\endbibitem

%%% 41
\bibitem[\protect\citeauthoryear{}{}]{gazebo}
\begin{botherref}
{Gazebo Simulator}.
\url{https://gazebosim.org/home}.
[Online]
\end{botherref}
\endbibitem

%%% 42
\bibitem[\protect\citeauthoryear{Sathaye et~al.}{2022}]{sathaye2022semperfi}
\begin{bchapter}
\bauthor{\bsnm{Sathaye}, \binits{H.}},
\bauthor{\bsnm{LaMountain}, \binits{G.}},
\bauthor{\bsnm{Closas}, \binits{P.}},
\bauthor{\bsnm{Ranganathan}, \binits{A.}}:
\bctitle{Semperfi: Anti-spoofing gps receiver for uavs}.
In: \bbtitle{Network and Distributed Systems Security (NDSS) Symposium 2022}
(\byear{2022})
\end{bchapter}
\endbibitem

%%% 43
\bibitem[\protect\citeauthoryear{Agyapong et~al.}{2021}]{agyapong2021efficient}
\begin{bchapter}
\bauthor{\bsnm{Agyapong}, \binits{R.A.}},
\bauthor{\bsnm{Nabil}, \binits{M.}},
\bauthor{\bsnm{Nuhu}, \binits{A.-R.}},
\bauthor{\bsnm{Rasul}, \binits{M.I.}},
\bauthor{\bsnm{Homaifar}, \binits{A.}}:
\bctitle{Efficient detection of gps spoofing attacks on unmanned aerial vehicles using deep learning}.
In: \bbtitle{2021 IEEE Symposium Series on Computational Intelligence (SSCI)},
pp. \bfpage{01}--\blpage{08}
(\byear{2021}).
\bcomment{IEEE}
\end{bchapter}
\endbibitem

%%% 44
\bibitem[\protect\citeauthoryear{Jullian et~al.}{2021}]{jullian2021deep}
\begin{bchapter}
\bauthor{\bsnm{Jullian}, \binits{O.}},
\bauthor{\bsnm{Otero}, \binits{B.}},
\bauthor{\bsnm{Stojilovi{\'c}}, \binits{M.}},
\bauthor{\bsnm{Costa}, \binits{J.J.}},
\bauthor{\bsnm{Verd{\'u}}, \binits{J.}},
\bauthor{\bsnm{Pajuelo}, \binits{M.A.}}:
\bctitle{Deep learning detection of gps spoofing}.
In: \bbtitle{International Conference on Machine Learning, Optimization, and Data Science},
pp. \bfpage{527}--\blpage{540}
(\byear{2021}).
\bcomment{Springer}
\end{bchapter}
\endbibitem

\end{thebibliography}
\end{document}